\documentclass[a4paper,11pt]{article}
\pdfoutput=1 

\usepackage{jcappub} 
\usepackage[T1]{fontenc} 

\usepackage{graphicx,amssymb,amsmath,amsthm,amsfonts,epsfig,times,bm}
\usepackage{aas_macros}
\usepackage{cleveref}

\usepackage{amsmath,amssymb}
\usepackage{tensor}

\def\bea{\begin{eqnarray}}
\def\eea{\end{eqnarray}}
\def\bes{\begin{subequations}}
\def\ees{\end{subequations}}

\title{Constraining coupled quintessence with the 21cm signal}

\author[a,b,c]{Xue-Wen Liu,}\emailAdd{liuxuewen14@itp.ac.cn}
\author[d]{Caroline Heneka,}\emailAdd{caroline.heneka@sns.it}
\author[c]{Luca Amendola}\emailAdd{l.amendola@thphys.uni-heidelberg.de}

\affiliation[a]{CAS Key Laboratory of Theoretical Physics,
Institute of Theoretical Physics, Chinese Academy of Sciences,
Beijing 100190, China}
\affiliation[b]{School of Physical Sciences, University of Chinese Academy of Sciences,
No. 19A Yuquan Road, Beijing 100049, China
}
\affiliation[c]{Institut f\"ur Theoretische Physik, Ruprecht-Karls-Universit\"at Heidelberg, \\ Philosophenweg
16, 69120 Heidelberg, Germany}

\affiliation[d]{Scuola Normale Superiore, Piazza dei Cavalieri 7, 56126 Pisa, Italy}

\abstract
{
The 21cm line probes the evolution of matter perturbations over a wide range of redshifts, from the dark ages to the completion of reionization, and down to the present-day Universe. Observing the 21cm cosmological signal will extend our understanding of the evolution of the Universe and it is thus important to investigate the predictions of different cosmological models. In this paper we focus on the prospect of constraining coupled quintessence models during the Epoch of Reionization both for  global signal experiments and   for intensity mapping  surveys. To derive the all-sky 21cm signal and  fluctuations in coupled quintessence, we simulate cosmological volumes of the 21cm signal including the coupling between dark matter and the quintessence field, where the strength of the coupling is labeled by the parameter $Q$. We show that the coupling between dark matter and quintessence modifies structure formation and   expedites the process of reionization. For upcoming 21cm line surveys like SKA and a fiducial global 21cm signal experiment, we perform a Fisher matrix analysis to constrain the coupling  $Q$ and the dark matter density parameter $\Omega_\mathrm{dm}$.
The results indicate that SKA will be able to place a 68\% upper limit of $0.04$ on $|Q|$.
At the same time, our fiducial global 21cm detector constrains the dark matter density parameter $\Omega_\mathrm{dm}$ with a predicted error of $\Delta\Omega_{\rm dm}\approx 0.005$, whereas SKA sets a weaker constraint of $\Delta\Omega_{\rm dm}\approx0.1$. These constraints are comparable to those already obtained from the cosmic microwave background, but explore an entirely different redshift range.
}

\begin{document}

\maketitle
\flushbottom

\section{Introduction}
\label{sec:intro}

Understanding the  late-time accelerated expansion of the Universe is one of the major challenges in cosmology~\citep{Riess:1998cb,Perlmutter:1998hx}. In the standard $\Lambda$CDM model, where gravity is described by General Relativity (GR), expansion is driven by a cosmological constant. Alternatively, for example in an extended $w$CDM model, the accelerated expansion is sourced by a form of energy with negative pressure different from a constant $w=-1$ in $\Lambda$CDM, called dark energy. In both models, cold dark matter does not interact with other energy components except through gravity.  Although these non-interacting scenarios are simple and successful in modeling the observed cosmic expansion, the true nature of the dark sector insofar it concerns the coupling or interaction of its possible components - dark matter and dark energy - is still to be determined. One motivation to consider the physics beyond the standard scenario  stems from the coincidence problem~\citep{Amendola:2000uh,Copeland:2006wr}. The observation that the energy densities of dark energy and dark matter are of the same order in the present Universe might point to a relation between these two  components. In combination with the notion of dark energy being described by a dynamic quintessence field $\phi$, the interaction between the dark matter fluid and the $\phi$-field has been therefore widely investigated in the literature (see for instance ~\citep{Wetterich:1994bg,Amendola:1999er, Chimento:2003iea}).

 An interaction between dark matter and the quintessence field~$\phi$ can influence the cosmic evolution at different scales. The coupled quintessence model employed here approximates at the background level the standard $\Lambda$CDM late-time cosmic accelerated expansion. The perturbative equations of motion at the linear level, however, are modified as compared to   $\Lambda$CDM. The dynamics of coupled quintessence models has been constrained by cosmic microwave background (CMB) observations at a redshift of $z\sim 1100$~\citep{Amendola:2000ub,Guo:2007zk,Olivares:2005tb,Wang:2006qw,Pettorino_2013}.  After CMB recombination the so-called dark ages follow, where the medium is cold and neutral, and structures continue to form and collapse, until the cosmic dawn followed by the Epoch of Reionization (EoR) when the first stars and galaxies arise and then ionise again the  neutral medium at redshifts of roughly $6< z < 10$. During all epochs, an interaction between dark matter and the quintessence field  leads to an extra attractive force,  so that structure formation is sped up. Therefore probes of the dark ages and the EoR provide additional constraints on the interaction between dark matter and the quintessence field at high redshifts beyond 6. While the 21cm line also probes structure at later times~\cite{2009MNRAS.397.1926W,xu2018constraining}, it is an especially promising probe of the EoR and beyond.

Upcoming and ongoing experiments like the Square Kilometre Array\footnote{https://skatelescope.org/}  (SKA)~\citep{2015aska.confE...1K} and the Experiment to Detect the Global EoR Signature (EDGES)~\citep{Bowman:2012hf} will constrain astrophysics and cosmology  with both power spectra of 21cm fluctuations and the global signal at redshifts of reionization and beyond, well into the dark ages.
During the heating epoch preceding reionization, the first generation of astrophysical sources emits both Lyman-$\alpha$ photons and X-ray radiation. At lower emissivities Lyman-$\alpha$ coupling occurs first between the cold gas and the spin temperature of hydrogen atoms, resulting in an absorption feature.  Increasing X-ray temperature later heats the intergalactic medium (IGM) above the CMB temperature, so that neutral hydrogen is seen in 21cm emission. The neutral hydrogen tends to trace the underlying density field at early epochs, with peaks in the density field resulting in peaks of 21cm emission. Later on, the first ionising sources start to ionise the medium around them, leaving the highest peaks in density void of 21cm emission that traces the neutral medium. Therefore the power spectrum of 21cm fluctuations folds in both the evolution of the matter distribution as well as the progress of reionization. In addition, the measurement of the global 21cm signal is complementary to measuring fluctuations, since it contains rich information about the formation and properties of the first galaxies. All in all, global 21cm and power spectrum measurements are indicative of the evolution of structures during the dark ages and the EoR, and thus sensitive to the coupling between dark matter and a possible quintessence field.

The prospect of constraining theories beyond $\Lambda$CDM with 21cm intensity mapping experiments has been investigated in~\citep{Hall:2012wd, Brax:2012cr}, and in~\citep{Heneka:2018ins} for general modifications to gravity.
The setup investigated here is similar to the one in ~\citep{Heneka:2018ins}, but differs in a constant coupling between dark matter and dark energy as compared to a general modification of the gravitational strength. Two further major differences should be stressed. First, in this paper we explore the constraints arising not just from the 21cm fluctuation power spectrum but also from the global 21cm signal (averaged over the full sky). As is well known, a first tentative detection of the global 21cm absorption signal during the dark ages was recently reported by the EDGES experiment \citep{Bowman:2018yin}. Secondly, contrary to the previous paper, we do not fix the background to $\Lambda$CDM or $w$CDM, but we evolve it self-consistently within the same coupled model. This requires a trial-and-error search for the initial condition of the cosmological expansion that give the present-day values of the cosmological parameters.

In section \ref{sec:cqev} we present the dynamics of coupled quintessence. In section \ref{sec:21cmcq} we use a semi-numerical simulation to calculate both the 21cm global signal in the redshift range $15<z<20$ and the 21cm power spectrum in the range $6<z<11$. The Fisher analysis for the global 21cm signal and 21cm power spectrum is shown in section \ref{sec:Fisher21cm}, and we present our conclusions in section \ref{sec:conc}.

\section{Coupled quintessence model}\label{sec:cqev}

In this section we briefly introduce the dynamics of the cosmic components in our interacting dark energy model. Within the coupled scenario, the energy-momentum tensors of quintessence (subscript $\phi$) and dark matter (subscript $\mathrm{dm}$) are not separately conserved; instead, the interaction leads to an energy exchange
\bea
 \nabla_{\mu}T^{\mu}_{\nu(\phi)}= - I_\mathrm{interaction},~~~ \nabla_{\mu}T^{\mu}_{\nu(\mathrm{dm})}= + I_\mathrm{interaction},
\label{eqtmnu}
\eea
where $T^{\mu}_{\nu(\phi)}$ and $T^{\mu}_{\nu(\mathrm{dm})}$ are the energy-momentum tensors of quintessence $\phi$ and dark matter, respectively.  In this paper we discuss  the interaction which arises because of a conformal coupling, and the term $I_\mathrm{interaction}$ has the form
\bea
I_\mathrm{interaction}= Q T_\mathrm{dm} \nabla_{\nu}\phi,
\eea
where $T_\mathrm{dm}$ is the trace of the energy-momentum tensor of the dark matter fluid and $Q$ is a parameter, indicating the strength of interaction between dark matter and dark energy. For simplicity, we shall use  units such that $8\pi G=1$, unless stated otherwise, and $\phi$ is taken in units of the Planck mass.

The Lagrangian density of the coupled scalar field is
\bea
\mathcal{L}_{\phi} = -(1/2)g^{\mu\nu}\partial_{\mu}\phi\partial_{\nu}\phi - V(\phi) + \mathcal{L}_{int},
\eea
where the interaction term $ \mathcal{L}_{int}$ gives rise to the interacting energy-momentum tensor given in Eq.~\eqref{eqtmnu}. For the potential we assume the exponential form,
\bea
V(\phi) = V_0 e^{-\lambda \phi},
\eea
where $\lambda$ is a dimensionless constant which can be assumed to be positive without loss of generality.
The parameter $\lambda$ gives the asymptotic value of the dark energy equation of state, $w_{\phi}\to -1+\lambda^2/3$.
In the following we choose both $\lambda=1$ and $\lambda=0.1$ to explore its effect on the forecasts.
Recent observational constraints found best-fitting values for $\lambda$ in the range $\left[0.3, 1.4 \right]$ in the coupled scenario~\cite{vandeBruck:2017idm,2019MNRAS.489..297A,Gomez-Valent:2020mqn} and ranging between $\left[0.1, 1.0 \right]$ for quintessence without coupling~\cite{2018arXiv180409350S}, depending on the combination of data sets. The normalization constant $V_0$ is fixed by estimating $\phi_0\sim M_\mathrm{P}$, where $M_\mathrm{P}$ is the Planck mass, and $\rho_\mathrm{de,0}\sim V\left( \phi_0\right)$ at present time, to get $V_0\sim \rho_\mathrm{de,0} \times e$.

In the coupled quintessence model, the quintessence field $\phi$ is thought to drive the late-time cosmic expansion, while dark matter  constitutes approximately 25\% of the total energy budget at present time in the Universe. Assuming a spatially flat Friedmann-Lema\^itre-Robertson-Walker metric, specified by the line element $ds^2 = dt^2 + a^2\delta_{ij}dx^idx^j$, the evolution of scalar field $\phi$, dark matter, baryons (subscript b) and radiation (subscript r) is governed by the following equations of motion, respectively,
\bes
\label{evolII}
\bea
{\ddot \phi} + 3 H {\dot \phi} + V'(\phi) &=& - Q \rho_\mathrm{dm} ,
\label{evol3}
\\\
\dot{\rho}_\mathrm{dm} + 3H  \rho_\mathrm{dm} &=&  Q \rho_\mathrm{dm} \dot{\phi},\label{evol2}
\\
\dot{\rho}_\mathrm{b} + 3 H  \rho_\mathrm{b} &=&  0,\label{evol2a}
\\
\dot{\rho}_\mathrm{r} + 4 H  \rho_\mathrm{r} &=&  0,\label{evol2b}
\eea
\ees
where a dot represents the derivative with respect to cosmic time and a prime to the scalar field $\phi$. The background evolution of the Universe is described by the Friedmann equation
\bea
3H^2 = \frac{\dot{\phi^2}}{2} + V(\phi) + \rho_\mathrm{dm} + \rho_\mathrm{b} + \rho_\mathrm{r}.
\label{H2}
\eea

It turns out to be convenient to use  the $e$-folding number $N=\log a$ as time variable and to introduce the following dimensionless variables,
\bea
x_1  \equiv  \frac{\dot{\phi}}{\sqrt{6}H},~~x_2 \equiv  \frac{\sqrt{V}}{\sqrt{3}H},~~ x_3 \equiv  \frac{\sqrt{\rho_\mathrm{r}}}{\sqrt{3}H},~~x_4 \equiv  \frac{\sqrt{\rho_\mathrm{b}}}{\sqrt{3}H}.
\eea
Eqs. \eqref{evolII} become then~\citep{amendola2010dark}
\bes
\label{evolnew}
\bea
\frac{dx_1}{dN}&=&-3 x_1 + \frac{\sqrt{6}}{2}\lambda x^2_2 - \frac{x_1}{H}\frac{dH}{dN}-\frac{\sqrt{6}}{2}Q  \nonumber\\
&~&\times \left(1-x^2_1-x^2_2-x^2_3-x^2_4\right),
\label{evolx1}
\\\
\frac{dx_2}{dN}&=& -\frac{\sqrt{6}}{2}\lambda x_1x_2 -  \frac{x_2}{H}\frac{dH}{dN},\label{evolx2}
\\
\frac{dx_3}{dN}&=& -2x_3 -  \frac{x_3}{H}\frac{dH}{dN},\label{evolx3}
\\
\frac{dx_4}{dN}&=& -\frac{3}{2}x_4 -  \frac{x_4}{H}\frac{dH}{dN}.\label{evolx4}
\eea
\ees
Moreover, taking the time derivative of Eq. \eqref{H2} in terms of e-folding $N$ and combining it with Eq. \eqref{evolII}, we obtain
\bea
  \frac{1}{H}\frac{dH}{dN}= -\frac{1}{2}(3 + 3x^2_1 - 3x^2_2 + x^2_3).
\label{H2xx}
\eea

\begin{figure}
\begin{center}
  \includegraphics[width=0.8\textwidth]{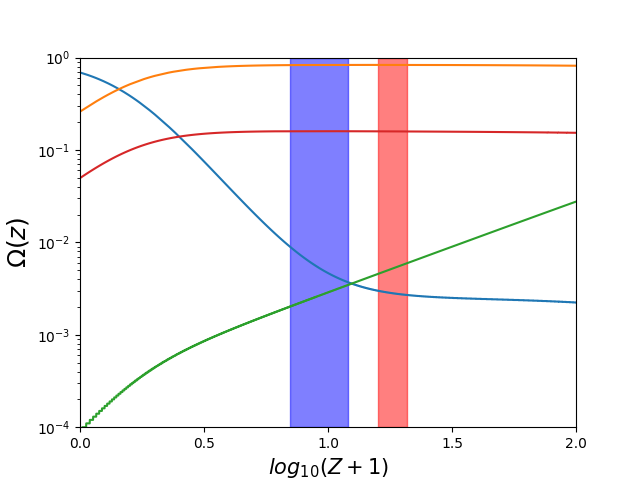}
  \caption{Evolution of cosmological density parameters in a coupled quintessence model with $Q=0.0735$. Blue line for the quintessence field, green line for the radiation field, red line for baryons and orange line for dark matter (bottom to top at high-$z$). The blue shaded region is the redshift range used for our Fisher analysis of the 21cm power spectrum and the red shaded region of the global 21cm signal.}\label{fig:cpOmgdz}
\end{center}
\end{figure}

\noindent The total effective equation of state of the universe is then
\bea
w_{\rm eff}= x^2_1 - x^2_2 + x^2_3/3.
\eea
The equation of state $w_{\phi}$ and the density parameter $\Omega_{\phi}$ for the scalar field are
\bea
w_{\phi}= \frac{x^2_1 -  x^2_2}{x^2_1 +  x^2_2},~~~\Omega_{\phi} =x^2_1 +  x^2_2.
\eea
The time-dependent density parameters can then be rewritten as
\bea
\Omega_\mathrm{r} = x_3^2,~~\Omega_\mathrm{b} = x^2_4,~~\Omega_\mathrm{dm} = 1 - x^2_1 - x^2_2 - x^2_3 - x^2_4.
\eea
The evolution of dark matter density perturbations can be obtained by perturbing the equations of motion Eqs. \eqref{evolII}. In the quasi-static regime, i.e. for sub-sound horizon scales, one finds the following equation
\bea
\frac{d^2\delta_\mathrm{dm}}{dN^2} &+& \frac{1}{2}(1-3 w_{\rm eff}+2\sqrt{6} Q x_1)\frac{d\delta_\mathrm{dm}}{dN} \nonumber\\
 &-& \frac{3}{2}(1+2Q^2) \delta_\mathrm{dm} \Omega_\mathrm{dm}=0.
\label{eomdc}
\eea
The  $Q^2$ term is independent of the sign of $Q$, while the term $Qx_1$ in the coefficient of $d\delta_\mathrm{dm}/dN$ is also proportional to $Q^2$ during the matter dominated era, and  usually very small with respect to the others. However, $Q$ also enters in  the background behavior, so the problem is not strictly symmetric around $Q=0$.
We also notice that $\lambda$ does not enter this equation, and therefore we expect the results of the 21cm power spectrum will be relatively independent of $\lambda$.

\begin{figure*}
  \includegraphics[width=0.5\textwidth]{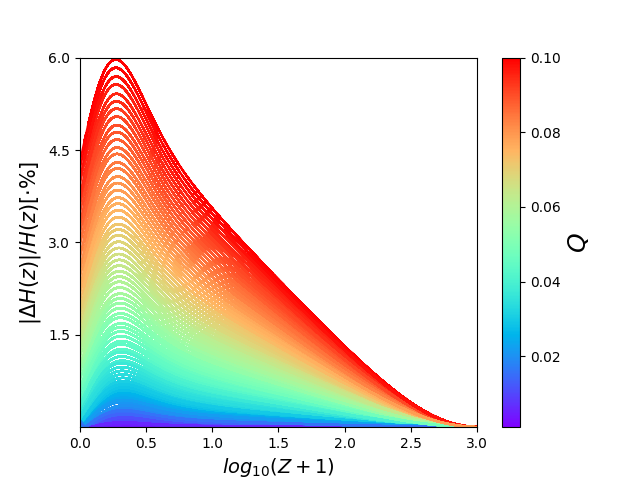}
  \includegraphics[width=0.5\textwidth]{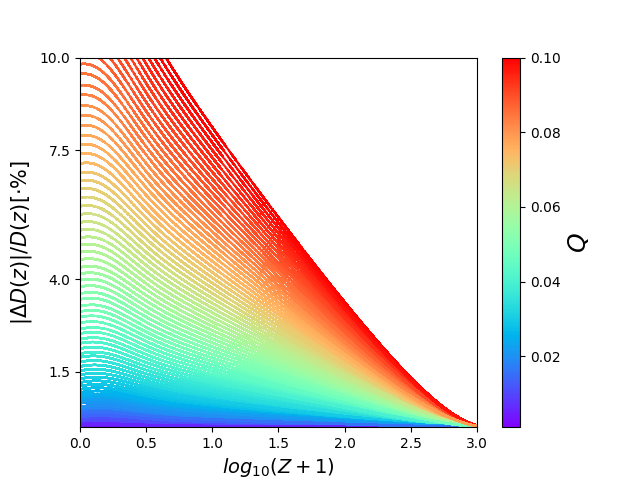}
  \caption{Hubble parameter (left panel) and dark matter growth factor  (right panel) in the coupled quintessence scenario with an exponential potential  $\lambda=1$ and  for different values of the coupling $Q$. }\label{fig:couplingHdz}
\end{figure*}

Unless stated otherwise, the fiducial value of the present cosmological parameters are chosen as: $\sigma_8=0.8, h=0.678, \Omega_\mathrm{r} = 8.6 \times 10^{-5}$ (including neutrinos), $\Omega_\mathrm{dm}=0.256, \Omega_\mathrm{b}=0.052$ ~\citep{Ade:2015xua}. In Figure~\ref{fig:cpOmgdz} we illustrate the typical evolution of cosmic energy components in our interacting quintessence model. The shaded regions are the redshift ranges used later in the Fisher matrix analysis. In contrast to  the standard cosmological scenario, the dark energy field contributes an almost constant $\approx 0.3\%$ percent to the energy budget between redshift 10 to 20 (more exactly, a fraction $\Omega_{\phi}=2Q^2/3$). Figure~\ref{fig:couplingHdz} shows the Hubble parameter (left panel) and the growth function for dark matter (right panel) when varying the coupling constant $Q$.  To show the impact of coupling between dark matter and the scalar field on the evolution of the Universe at present time, we have chosen to normalize the curves at the CMB redshift $z=1100$. The range of  $Q$ is chosen to be within the $2\sigma$ confidence bounds obtained for ${\rm CMB}+{\rm SNIa}$ data~\citep{Xia:2013nua}.  The deviation of the Hubble parameter with respect to $\Lambda$CDM increases with  the coupling and reaches a maximal value of 6$\% $ shortly before the scalar field dominates the Universe at present time. Similarly, the deviation of the growth function also increases with stronger couplings. The deviation from the standard case, however, keeps growing until today, and reaches a maximal $\mathcal{O}(10\%)$ deviation. These substantial deviations from $\Lambda$CDM have a strong impact on the 21cm signal, as we will show in the following sections.

\section{The 21cm signal}\label{sec:21cmcq}

In this section, we present the signature in the global 21cm signal and power spectrum caused by the coupling in the dark sector. The 21cm signal can be recorded in absorption or emission relative to the background CMB temperature with a characteristic rest frame wavelength of $\lambda_0=$21cm and corresponds to the forbidden spin-flip transition of neutral hydrogen.  It depends strongly on the fraction and density of neutral hydrogen, as well as  on the so-called spin temperature (in particular during the heating epoch that precedes reionization). The spin temperature is determined by the ratio of Boltzmann levels for the forbidden spin-flip transition of
neutral hydrogen in its ground state, i.e. the relative number of hydrogen atoms with proton and electron spin aligned or anti-aligned.

The 21cm brightness temperature is expressed in a general cosmology as the differential brightness temperature relative to the CMB temperature as~\citep{2006PhR...433..181F}
\begin{align}
\delta T_\mathrm{b} \left({\bf x}, z \right) &= \frac{T_\mathrm{S}-T_{\gamma}}{1+z}\left(1- e^{-\tau_{\nu_0}} \right) \nonumber \\
& \approx  27x_\mathrm{HI}\left( 1+\delta_\mathrm{nl}\right)\left( \frac{E(z)}{\mathrm{d}v_\mathrm{r}/\mathrm{d}r + E(z)}\right)\left( 1- \frac{T_{\gamma}}{T_\mathrm{S}}\right) \nonumber \\ & \vspace{0.2cm}\times (1+z)^2 \left(\frac{0.1225}{E(z)}\right)\left( \frac{\Omega_\mathrm{b} h^2}{0.023}\right) \mathrm{mK} , \label{eq:Tbgen}
\end{align}
where $T_{\gamma}$ is the CMB temperature, $\Omega_\mathrm{b}$ the present-day baryon density parameter,   $T_\mathrm{S}$ the gas spin temperature, $\tau_{\nu_0}$ the optical depth at the rest frame frequency $\nu_0$, $x_\mathrm{HI}$ is the hydrogen neutral fraction, $\delta_\mathrm{nl}\equiv \rho/\bar{\rho} - 1$ the evolved density contrast, $H(z)$ the Hubble parameter, which can be written as $H(z)=H_0E(z)$, and $dv_\mathrm{r}/\mathrm{d}r$ is the comoving gradient of the comoving velocity along the line of sight in units of $H_0$.

\begin{figure}
\begin{center}
  \includegraphics[width=0.8\textwidth]{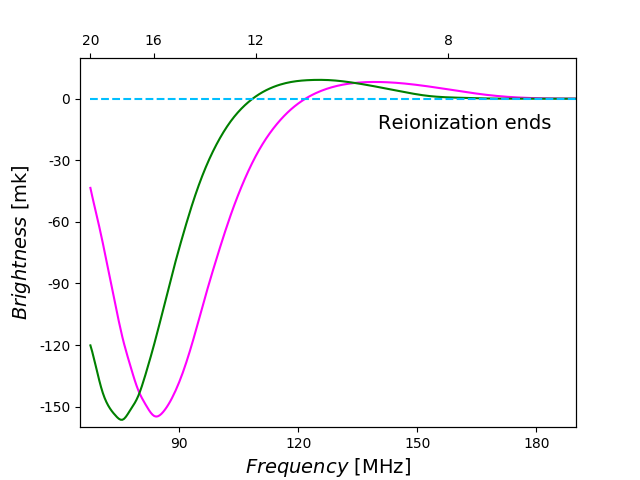}
  \caption{Global 21cm signal as a function of redshift for a coupled quintessence model with $Q=0.0735$ (green line) and for no coupling (violet line, note the shift to higher frequencies and lower redshifts, respectively) .}\label{fig:21cmglobal}
\end{center}
\end{figure}

The global signal is taken as the full-sky average of the 21cm signal and the 3D fluctuations are derived from the spatial variation in the radiation fields and the density, ionization and temperature states of the IGM. Fluctuations might arise from fluctuations in the baryon density, neutral fraction, Ly$\alpha$ coupling coefficient, gas temperature and line-of-sight peculiar velocity gradient. In practice we make use of a modified version of the semi-numerical code \texttt{21cmFAST}~\footnote{\url{http: //github.com/andreimesinger/21cmFAST}} to produce simulation boxes of 21cm fluctuations and derive the full-sky global signal at a given redshift. We incorporate the coupling between dark matter and quintessence fields at background and linear perturbation level to calculate the corresponding evolution of the Hubble parameter and the linear density field, and then use the Zel'dovich approximation to create the velocity field and to calculate the velocity gradient along the line of sight. The ionisation field is calculated by adopting an excursion set approach~\citep{Furlanetto:2004nh}.  Here the overdensities are first filtered at different radii~$R_i$, and the region with the radius~$R_i$, whose filtered overdensity is higher than a collapsed fraction set by an ionising efficiency $\zeta$, is assumed to be ionised.

We choose the fiducial reionization model parameters in the following calculation as
\begin{align}
T_\mathrm{vir} = 4 \times 10^4 K,~~~R_\mathrm{MFP} = 20 Mpc,~~~\zeta = 31.5,\label{Asfidvalue}
\end{align}
where $T_\mathrm{vir}$ is the typical halo viral temperature, $R_\mathrm{MFP}$ is the mean free path of ionising photons, and $\zeta$ the ionising efficiency. For the heating history, we choose
\begin{align}
\zeta_\mathrm{X} = 2 \times 10^{56},~~~f_* = 0.05 ,\label{Asfidvalue2}
\end{align}
with the X-ray heating efficiency $\zeta_\mathrm{X}$ as the number of X-ray photons per solar mass in stars, and the fraction of baryons converted to stars $f_*$. With our choice of fiducial parameters the optical depth for the $\Lambda$CDM model is derived as $\tau_\mathrm{fid}=0.061$, which is consistent with the CMB constraint given by Planck~\citep{Adam:2016hgk}.

The evolution of the 21cm brightness temperature $T_\mathrm{b}$ is thus driven by the evolution of $x_\mathrm{HI}$, $\delta_\mathrm{nl}$ and $T_\mathrm{S}$ and is illustrated in Figure~\ref{fig:21cmglobal} for redshifts $6<z<20$ in both coupled quintessence and a scenario without coupling. Generally, the shape of the absorption signal can be explained as follows. After the first generation of astrophysical sources turned on around $z\sim 30$, these sources start to emit Ly$\alpha$ photons, which resonantly scatter off the hydrogen atoms, coupling the spin temperature $T_\mathrm{S}$ to the gas temperature $T_\mathrm{K}$ via the Wouthuysen-Field effect. Meanwhile, X-ray heating is not strong enough at this epoch to stop cooling and the kinetic temperature from decreasing. The Ly$\alpha$ coupling begins to saturate as the spin temperature couples to the kinetic temperature in most regions of the IGM. This produces a (relatively sharp) minimum close to $z\sim 16$ in our fiducial model. Increasingly strong X-ray emission heats up the gas more efficiently than Ly$\alpha$ emission before, so that the gas temperature is increased up to the CMB temperature $T_{\gamma}$ and beyond. This leads to an emission signal, as the gas continues to be heated to temperatures $T_\mathrm{K} > T_{\gamma}$, reaching during reionization the so-called post-heating regime when $T_\mathrm{S} \gg T_{\gamma}$ via coupling to $T_\mathrm{K}$. Thus during the EoR and the Dark Ages the two main contributions to the spin temperature are the coupling to the Ly$\alpha$ background via Wouthuysen-Field effect and then the collisional coupling to the kinetic gas temperature; we used the implementation of the spin temperature calculation as described in~\cite{2011MNRAS.411..955M}. Finally, reionization occurs as UV photons produce bubbles of ionised regions that permeate the Universe more and more, wiping out the 21cm signal inside ionised regions and leading to an overall decrease towards zero of the global 21cm signal at the end of reionization at $z \sim 6$.

For the forecast of constraints during the cosmic dawn and EoR in the following section, we used two approaches depending on the epoch treated. For constraints that employ the 21cm power spectrum during the EoR in section~\ref{sec:FisherPk}, we worked in the post-heating regime of $T_\mathrm{S}\gg T_{\gamma}$, approximating $\left( 1-T_{\gamma}/T_\mathrm{S}\right)\sim1$ in Eq. \eqref{eq:Tbgen}. During the dark ages at higher redshifts for lower $T_\mathrm{S}$ values one needs to follow the coupling to the kinetic gas temperature and the Ly$\alpha$ background to calculate the spin temperature field and then the 21cm brightness temperature, which we did for the forecast that employes the global 21cm signal in section~\ref{sec:Fisherglobal}. Typical spin temperatures in our model beyond redshift 12 evolve around $6$mK and less.

The interesting difference between this coupled  model and the standard uncoupled scenario is  that the coupling leads to an enhanced pull on dark matter, such that the dark matter perturbations grow quicker than without coupling. This leads to an earlier generation of the first  astrophysical sources, and consequently to an earlier production of  Ly$\alpha$ and  X-ray emission, assuming the same astrophysical settings. This induces  the behaviour displayed in Figure~\ref{fig:21cmglobal}, where the amplitude of minimal absorption and maximal emission for the global 21cm signal stay the same, but are shifted in redshift.

\begin{figure*}
  \includegraphics[width=0.50\textwidth]{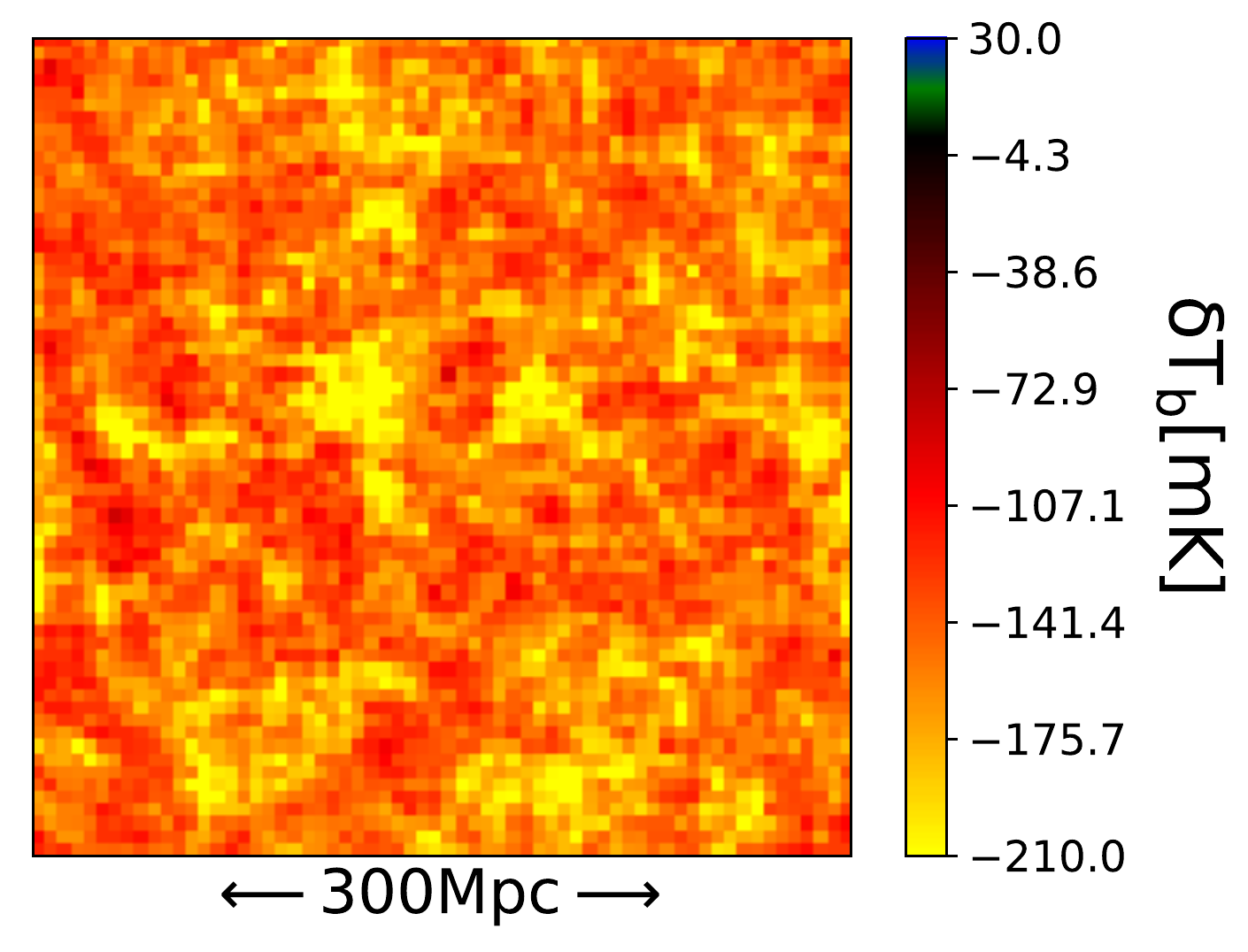}
  \includegraphics[width=0.50\textwidth]{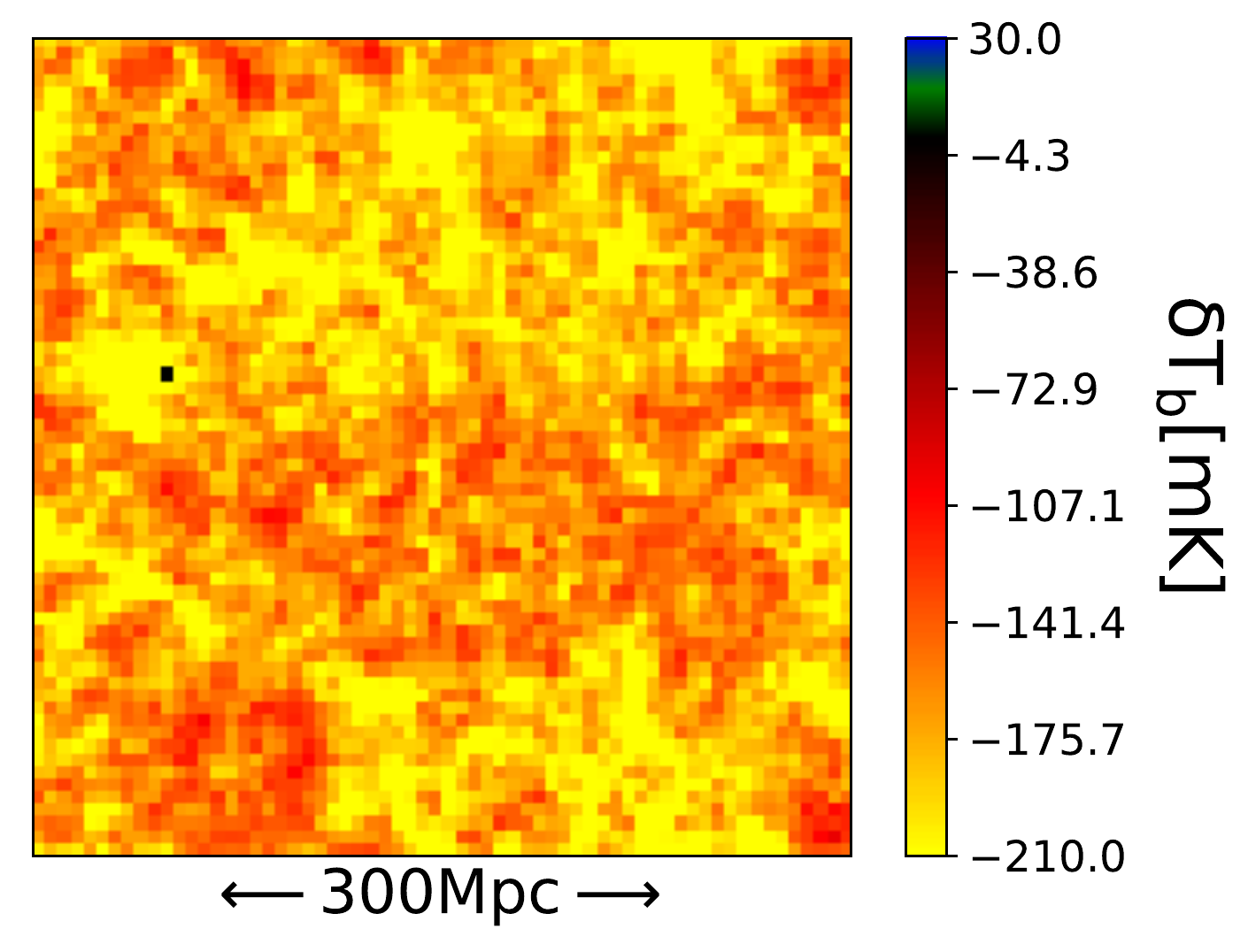}\\
  \includegraphics[width=0.50\textwidth]{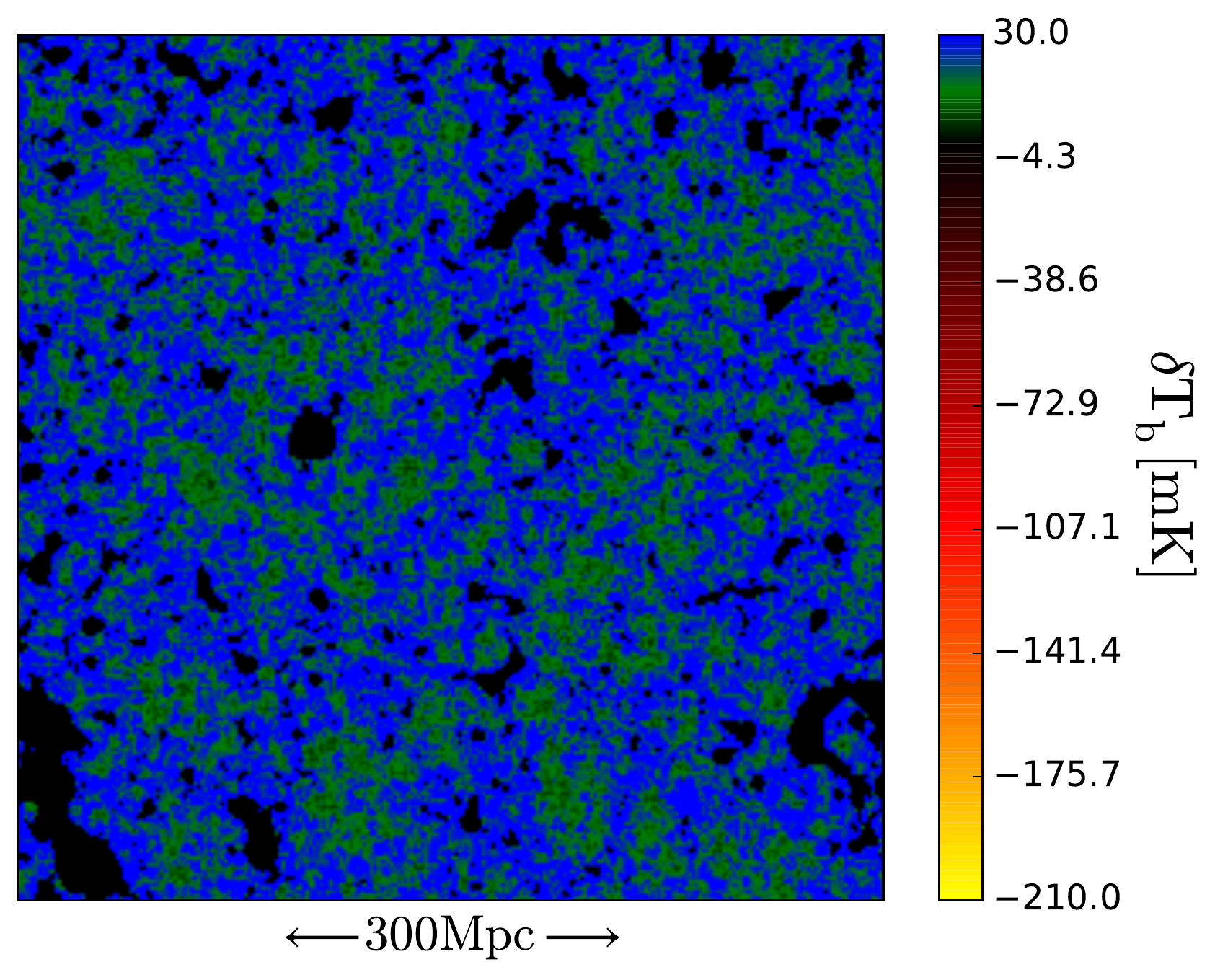}
  \includegraphics[width=0.50\textwidth]{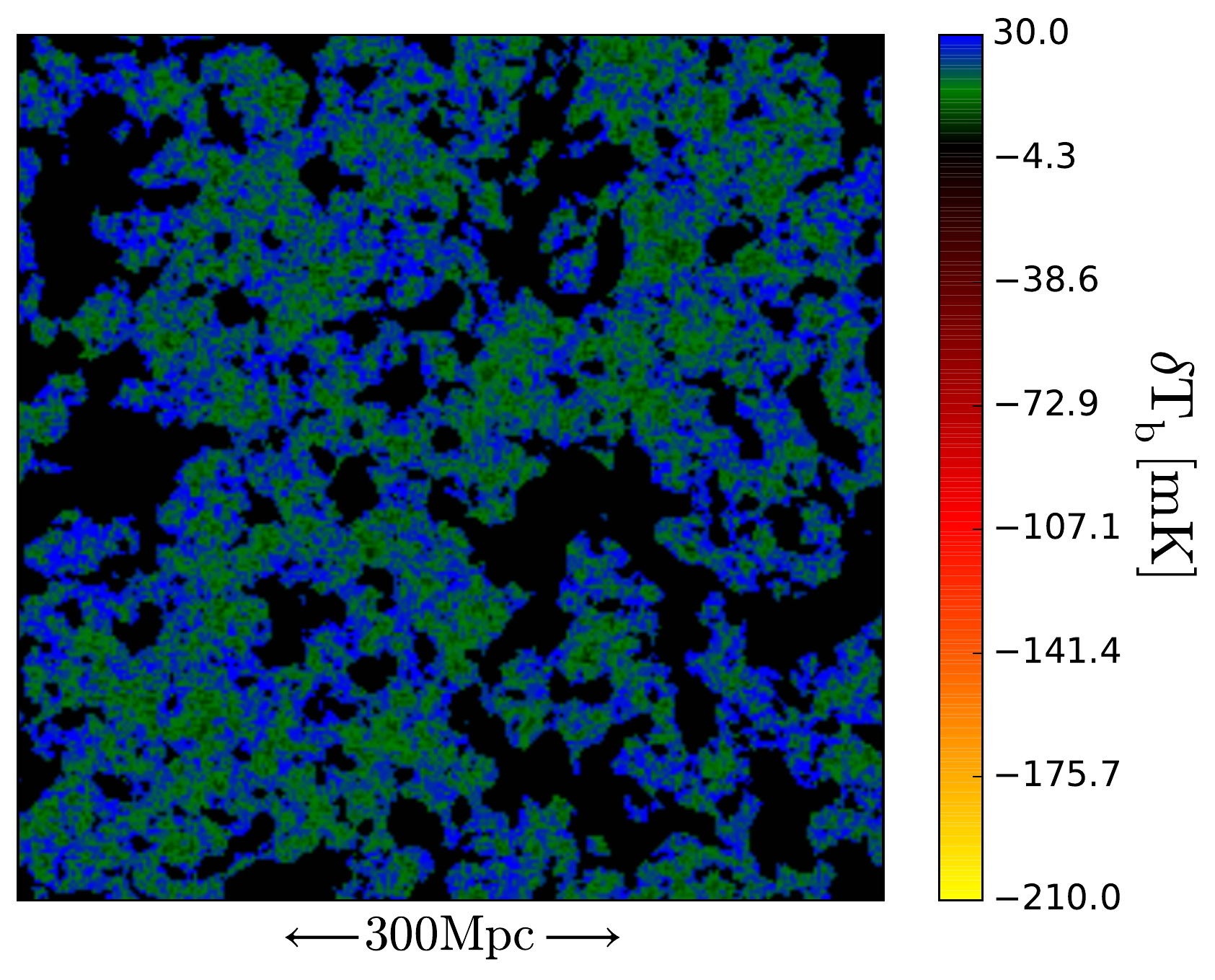}\\
  \includegraphics[width=0.50\textwidth]{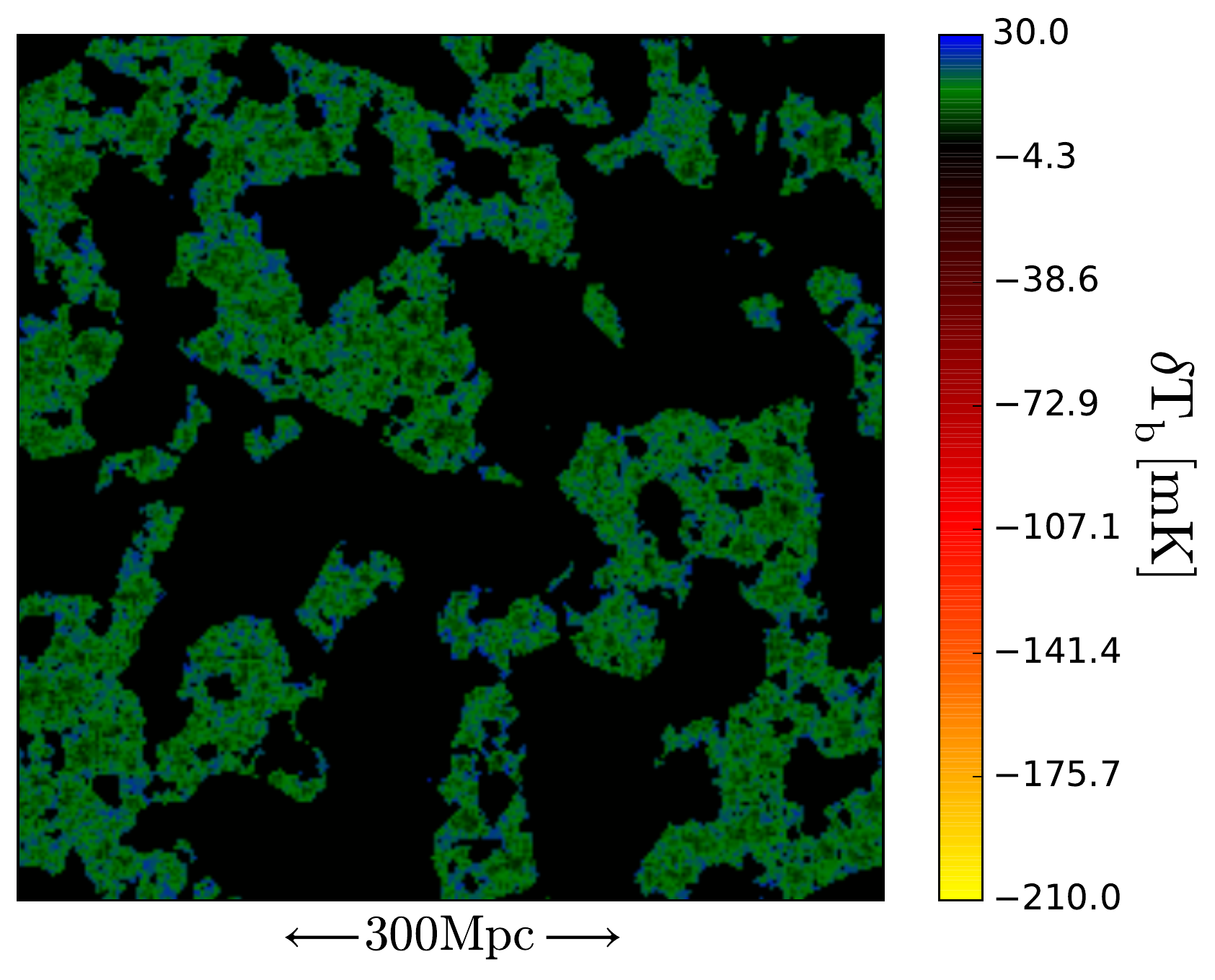}
  \includegraphics[width=0.50\textwidth]{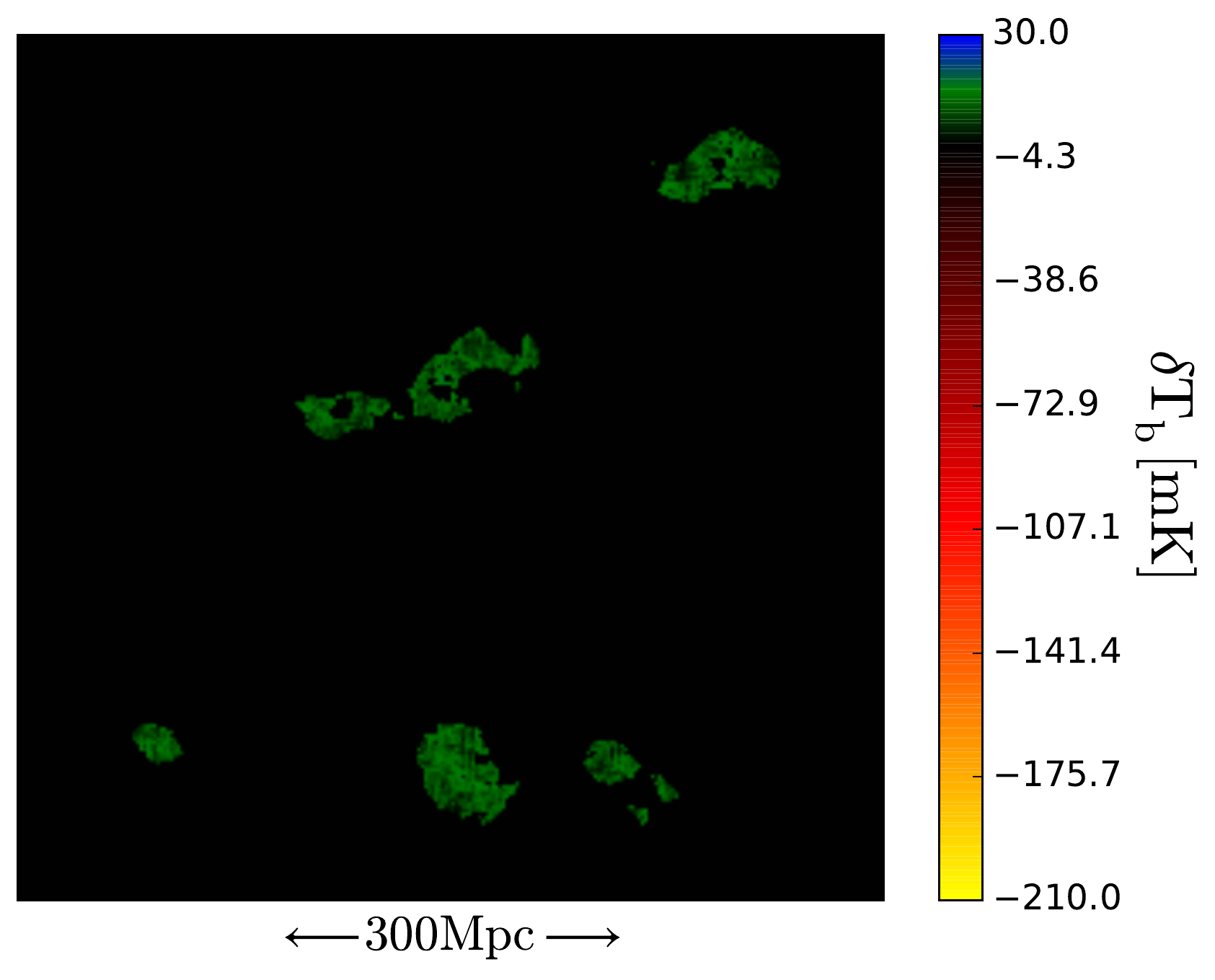}
  \caption{Comparison of 21cm emission at redshift $z=16$ (upper panels), $z=10$ (middle panels) and $z=7$ (lower panels) between the $\Lambda$CDM model (left panels) with $w=-1,\Omega_\mathrm{m}=0.308$ and the coupled quintessence model (right panels) with $\lambda=1.0$, $Q=0.0735$. Simulation box size is 300$\,$Mpc, resolution is 4.69$\,$Mpc for the global signal calculation between $z=20$ to $z=15$ and 1.17$\,$Mpc for power spectrum calculation between $z=11$ to $z=6$, see section~\ref{sec:21cmcq} for a more detailed description.}\label{fig:21cmZ10}
\end{figure*}

For both the coupled quintessence model and the standard $\Lambda$CDM scenario we evolve our simulations down from high redshifts of $z\sim 20$ to the end of reionization at $z\sim 6$.
In Figure~\ref{fig:21cmZ10} we illustrate the spatial fluctuations of the 21cm brightness temperature both during the heating epoch and during reionization at redshifts $z=16$, $z=10$ and $z=7$ with simulation box size 300 Mpc. We can draw a similar  conclusion as for the global signal for the behaviour of the uncoupled (left panel) versus the coupled (right panel) model: with coupling between dark matter and the quintessence field, structure formation and reionization start earlier and have progressed further by redshift $z=10$ and $z=7$, so that the 21cm signal shown is  generally fainter compared to the standard scenario at these redshifts.

To derive cosmological information from the spatial 21cm fluctuations, besides the mean global signal, we use the power spectra derived from our  simulated 21cm line emission at different redshifts. To estimate the  ability to detect 21cm power spectra during the epoch of reionization, we calculate the sensitivity for the upcoming Square Kilometre Array (SKA) interferometer ~\citep{2015aska.confE...1K}. We define the dimensionless 21cm power spectrum as $\tilde{\Delta}_{21}(k) = k^3 / (2\pi^2V)<|\delta_{21}|^2>_k$ and the dimensional power spectrum as $\Delta_{21}(k)=\bar{T}^2_{21}\tilde{\Delta}_{21}(k) $.

For our error estimate we assume the survey specifics of an SKA-LOW like stage 1 intensity mapping survey. The variance for a 21cm power spectrum for angle $\mu$ between the line of sight and mode $k$ is~\citep{McQuinn:2005hk}
\begin{equation}
\sigma^2_{21} \left( k,\mu\right) = \left[  P_{21}\left( k,\mu\right) + \frac{T_\mathrm{sys}^2 V_\mathrm{sur} \lambda_{21}^2}{B\,t_\mathrm{int}n\left( k_{\perp}\right) A_\mathrm{e}}W_{21}\left(k,\mu \right) \right] , \label{eq:sigma21}
\end{equation}
where the first term accounts for cosmic variance, the second term for thermal instrumental noise, and the window function $W_{21} \left( k,\mu \right)$ accounts for the limited spectral and spatial instrumental resolution that translates to limiting windows in parallel and perpendicular modes. The effective survey volume takes the form  $V_\mathrm{sur}=\chi^2\Delta\chi\left( \lambda_{21}\left( z\right)^2/A_\mathrm{e}\right)$, at 21cm wavelength $\lambda_{21}\left(z \right)$, for an effective area $A_\mathrm{e}=925$m$^2$, and comoving distance and survey depth $\chi$ and $\Delta \chi$.
We take the spectral resolution $\nu_\mathrm{res}=3.9$kHz, a maximum baseline $l_\mathrm{max}=10^5$cm, an instrument system temperature of $T_\mathrm{sys}=400$K, a total observing time time of $t_\mathrm{int}=1000$hrs, and a survey bandwidth of $B=8$MHz~\citep{Chang:2015era, Pritchard:2015fia}.
The total variance $\sigma^2\left(k \right)$ of the spherically averaged power spectrum is to be summed over all angles $\mu$, and to be divided by the number of modes per bin; we explicitly counted the number of modes in each bin.
We caution though that assuming the number density of baselines $n_{\perp}$ to be constant can underestimate the error at higher $k$ values, as for longer baselines the number density decreases.
We also account for an angular size of roughly 100$\,$deg$^2$ and the corresponding survey volume.
For foreground treatment we chose foreground avoidance and cut away in $k$-space the so-called foreground wedge~\citep{Morales:2012kf}. The foreground wedge for the cylindrically averaged 2D power spectrum is given by
\begin{equation}
k_{\parallel}  \leq \frac{\chi(z)E(z)\theta_0}{d_\mathrm{H}(1+z)}k_{\perp},
\end{equation}
with  parallel and orthogonal modes $k_{\parallel} $ and $k_{\perp}$, comoving distance $\chi(z)$, the dimensionless Hubble parameter $E(z) = H(z)/H_0$ and Hubble distance $d_\mathrm{H}$, as well as characteristic angle $\theta_0$, which we set to $10^{\circ}$. This corresponds to the assumption that contamination from residual sources is limited to the primary beam of the instrument.

\begin{figure}
\begin{center}
  \includegraphics[width=0.8\textwidth]{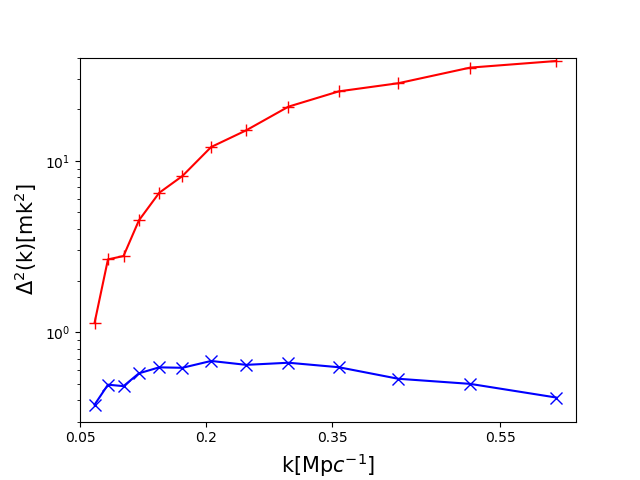}
  \caption{In red, 21cm power spectrum at redshift $z=10$ as a function of scale $k$; in blue, the line depicts the noise contribution.}\label{fig:21cmPowerSNR}
\end{center}
\end{figure}

In Figure~\ref{fig:21cmPowerSNR} we show the 21cm power spectrum as a function of scale $k$ at redshift $z=10$ together with the corresponding noise level. The noise is relatively flat with scale $k$, and slightly drops towards both low and high-$k$ values. Before it goes down drastically, the signal-to-noise for 21cm power spectrum measurements can reach values of $\mathcal{O}(10)$ in the range of scales depicted, where the $k$ range was limited to linear scales at these redshifts, $k<0.67$Mpc$^{-1}$.

\section{Fisher forecast} \label{sec:Fisher21cm}

To quantify the ability of future 21cm experiments like SKA to constrain the coupling $Q$ between dark matter and dark energy, we perform in this section forecasts  based on the Fisher matrix formalism.  On the cosmological side, we  vary both $Q$ and the dark matter density parameter $\Omega_\mathrm{dm}$. Moreover, since the 21cm signal strongly depends on the underlying astrophysics, we chose as astrophysical parameters $(T_\mathrm{vir}, R_\mathrm{MFP}, \zeta, \zeta_\mathrm{X}, f_*)$. Their fiducial values are given in Eq.~\eqref{Asfidvalue} and~\eqref{Asfidvalue2}. The fiducial set of cosmological parameters is taken as: $\sigma_8=0.8, \Omega_\mathrm{dm}=0.256, h=0.678, \Omega_\mathrm{b}=0.052, \Omega_\mathrm{r}=8.6\times 10^{-5}$. We forecast constraints for both the global 21cm signal and power spectrum because they test different redshift ranges, during  both the EoR and the heating epoch. Generally, global signal and power spectrum are sensitive to a different subset of parameters; for example  the global signal is sensitive to the heating-relevant parameters $T_\mathrm{vir}, \zeta_\mathrm{X}$, and  $f_*$ when the absorption signature at high redshifts during the epoch of heating is measured.

\subsection{Fisher analysis for the global 21cm signal}\label{sec:Fisherglobal}
The detection of the global 21cm signal is complementary to the power spectrum measurements, insofar as its high-redshift absorption feature during the heating epoch is sensitive to the properties of the first ionising sources. Examples for experiments that strive to measure the global 21cm signal include the Large Aperture Experiment to Detect the Dark Ages (LEDA)~\citep{price2018design}, the Long Wavelength Array (LWA)~\citep{2010iska.meetE..24H}, and the currently running Experiment to Detect the Global EoR Signature (EDGES)~\citep{Bowman:2007su}. For the global signal detection, on the one hand, one can use much less integration time to reach the required sensitivities compared with the power spectrum measurement. On the other hand, the foregrounds are spectrally smooth and the global signal is very sensitive to the redshift or frequency, which leads to the difficulty to separate the signal from foregrounds. The EDGES experiment reported an absorption profile at $\sim $ 78 MHz, with a width of 20 MHz, corresponding to a redshift range of $17<z<19$ with an amplitude of 0.5K~\citep{Bowman:2018yin}. If the signal is confirmed, this would be the first direct 21cm signal measurement from cosmic dawn. The anomalous amplitude and U-shaped absorption profile has caused large interest in non-standard interactions, for example between dark matter and baryons~\citep{Fialkov:2018xre,DAmico:2018sxd,Costa:2018aoy}. We therefore also take a closer look at the promising absorption signal during heating and focus on the redshift range $15\leq z \leq 20$ to derive constraints  on $Q$ for a general global 21cm experimental setup.

In general the Fisher matrix takes the form~\citep{1997ApJ...480...22T}
\begin{equation}
F_{ij}=\frac{1}{2}{\rm Tr}\left[C^{-1}C_{,i}C^{-1}C_{,j}+C^{-1}(\mu_{,i}\mu^T_{,j}+\mu_{,j}\mu^T_{,i})\right].
\end{equation}
where $C$ is the data covariance matrix, $C_{_,i}$ denotes the derivative with respect to the $i$-th parameter, and $\mu$ is the vector of the expectation values for the observable; the trace sums over the data.

In the case of a global 21cm experiment the observable is the antenna temperature $T_{\rm sky}$
\begin{equation}
T_{\rm sky}(\nu)=T_{\rm fg}(\nu)+T_\mathrm{b}(\nu),
\label{eq:Tsky}
\end{equation}
including the foreground temperature $T_{\rm fg}$ and the 21cm mean brightness temperature $T_\mathrm{b}$, all as a function of frequency $\nu$. By adopting a bandwidth $B$ we divide the signal into $N_{\rm ch}$ frequency bins ${\nu_n}$ in accordance with our full frequency or redshift range.  To account for the foreground contribution, notice that at the frequencies of interest, roughly 10--250 MHz, the sky is dominated by synchrotron emission from our Galaxy. We use a polynomial fitting form to describe the foreground temperature,
\bea
\log T_{\rm fg} =  \log T_0 + \sum^{N}_{i=1}a_i \log(\nu/\nu_0)^i,
\label{eq:sync}
\eea
where we adopt a 3rd order polynomial with pivot temperature $T_0=875 K$ and best-fit parameter values $ a_1 = -2.47, a_2 = -0.089, a_3=0.013$,  fitted to the frequency range $\nu = 50 - 150$MHz~\citep{Pritchard:2010pa}.

The detection in different frequency bins is expected to be uncorrelated, therefore the covariance matrix is taken to be diagonal
\begin{equation}
C_{ij}=\delta_{ij}\sigma_i^2.
\end{equation}
The noise $\sigma_i$ accounts for foreground residuals, limited resolution and thermal noise for a global 21cm signal detection and can be written as ~\citep{Liu:2012xy}
\bea
\sigma^2_i= T^2_\mathrm{sky}(\nu_i)\left(\frac{\epsilon^2_0\theta^2_\mathrm{fg}}{4\pi f_\mathrm{sky}} + \frac{1}{t_\mathrm{int}B}\right).
\label{eq:globalerr}
\eea
Here the parameter $\epsilon_0$ is used to model the residual foreground fraction in the signal and $\theta_{\rm fg}$ is the "native" angular resolution of the foreground model. Based on the results from~\citep{de2008model}, which found as a conservative estimate a $90\%$ accuracy at $\theta_\mathrm{fg}=5^{\circ}$ for the foreground cleaning, we therefore use $\epsilon_0=0.1$ and $\theta_\mathrm{fg}=5^{\circ}$ as fiducial foreground values. Furthermore, $f_\mathrm{sky}$ is the fraction of sky used in the experiment.  For a specific experiment, the factor $4\pi f_\mathrm{sky}$ is to be taken as the number of pixel times the solid angle of each pixel; here we choose $f_\mathrm{sky}=0.2$. Finally, $t_\mathrm{int}$ is the total integration time and $B$ is the band width. For these two parameters we take $B = 1\,$MHz and $t_\mathrm{int}=500$ hrs for our fiducial experimental setup.

Combining Eq.~\eqref{eq:Tsky} and~\eqref{eq:globalerr} leads to the following expression for the Fisher matrix of a global 21cm signal measurement
\begin{align}
F_{ij}=\sum_{n=1}^{N_\mathrm{ch}}\left(2+\left(\frac{\epsilon^2_0\theta^2_\mathrm{fg}}{4\pi f_\mathrm{sky}} +\frac{1}{t_\mathrm{int}B}\right)^{-1}\right) \frac{d\log T_{\rm sky}(\nu_n)}{d p_i} \frac{d\log T_{\rm sky}(\nu_n)}{d p_j},
\end{align}
where the parameter set $\{p_i\}$ includes both cosmological and astrophysical model parameters. For the global 21cm signal these parameters consist of \{$T_\mathrm{vir}, Q, \Omega_\mathrm{dm}, \zeta_\mathrm{X}, f_*$\}. From our full set of parameters, we keep both $\zeta$ and $R_\mathrm{MFP}$ fixed to their fiducial values, as the global signal during the early heating epoch is not sensitive, or only marginally sensitive, to the ionization efficiency $\zeta$ during reionization, and the mean free path of ionising radiation $R_\mathrm{MFP}$~\citep{Cohen:2016jbh}.
Since the fiducial value of $\zeta_\mathrm{X}$ is much larger than the other parameters,  we use $\log\left(\zeta_\mathrm{X}\right)$ as variable in our analysis instead of $\zeta_\mathrm{X}$ to improve numerical precision.

\begin{figure*}
  \includegraphics[width=1.0\textwidth]{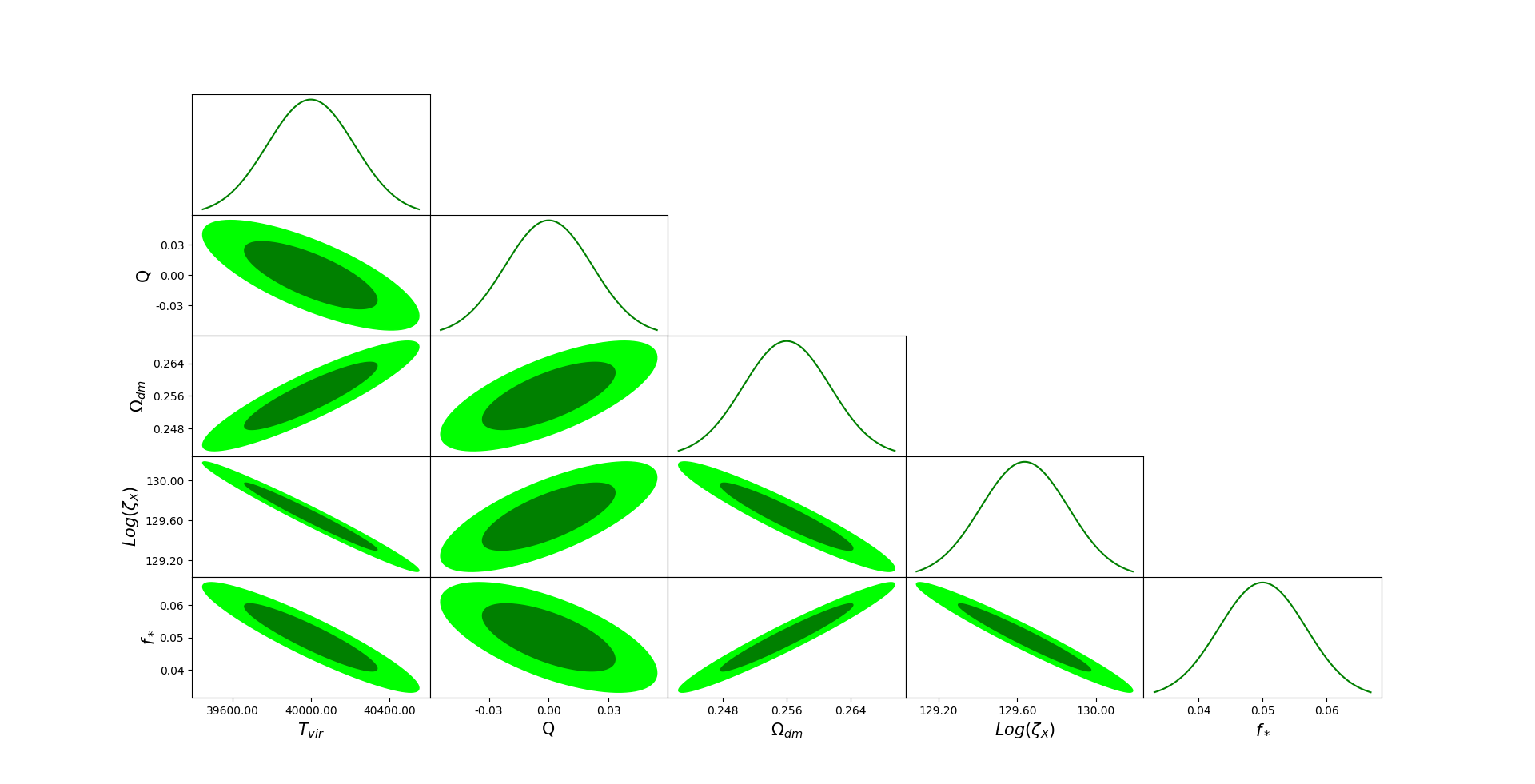}
  \caption{68.3\% and 95.4\% confidence ellipses derived by Fisher forecasts of  21cm global signal measurements, in the redshift range  from $z=20$ to $z=15$. The fiducial value for parameters ($T_\mathrm{vir}, Q, \Omega_\mathrm{dm}, \log \zeta_\mathrm{X}, f_*$) are set as ($4\times 10^4, 0.0, 0.256, 129.64, 0.05$) and $\lambda$ is chosen as 1.}\label{fig:2Dcounterback}
\end{figure*}

In Table~\ref{tab:1sigmapmsglob} we show for our fiducial global 21cm experiment the marginalised 1$\sigma$ confidence limits for our set of astrophysical and cosmological parameters and in Figure~\ref{fig:2Dcounterback} the corresponding 2D confidence contours. We can see that, by using the global 21cm signal, we can constrain the coupling constant $Q$  down to the $\mathcal{O}(0.02)$ level, which is close to the constraints achievable with CMB measurements~\citep{Xia:2013nua,Ade:2015rim}. Comparing the confidence limits for different $\lambda$ values, which vary the steepness of the quintessence potential,  we see in Table~\ref{tab:1sigmapmsglob} that the constraints are sensitive to the $\lambda$ fiducial value, but stay within one order of magnitude. We note that for a smaller value of $\lambda$, the corresponding constraints on the interaction $Q$ are found to be weaker, while the uncertainty on the fraction of baryons in stars $f_*$ will roughly stay the same with a couple of percent uncertainty at the 68\% confidence level. Moreover, the global 21cm signal provides better constraints on both the astrophysical parameters $T_\mathrm{vir}$ and $f_*$, as well as on the dark matter fraction $\Omega_\mathrm{dm}$ in comparison to the power spectrum results (see the next section).

\subsection{Fisher analysis for the 21cm power spectrum}\label{sec:FisherPk}
For the Fisher analysis of the 21cm power spectrum measurements we consider the SKA experiment~\citep{Santos:2015gra}, which aims at 21cm intensity mapping during the epoch of reionization. More explicitly,  we assume a SKA stage 1 configuration; the corresponding instrument specifications are listed in Table~\ref{tab:exp}. For our analysis we consider 6 redshift bins from $z=6$ to $z=11$ binned in steps of $\Delta z = 1$. For simplicity we make the assumption of working in the post-heating regime with $T_\mathrm{S} \gg T_{\gamma}$.

For the measurement of 21cm power spectra with an intensity mapping experiment the Fisher matrix is given by
\begin{align}
F_{ij} = \sum_{z,k} \frac{\Delta k k^2 V_\mathrm{sur}}{4\pi^2} \frac{\partial \tilde{\Delta}_{21}^2(z,k)}{\partial p_{i}} C^{-1}(z,k)  \frac{\partial \tilde{\Delta}_{21,l}^2(z,k)}{\partial p_j} ,
\end{align}
where $\tilde{\Delta}_{21}$ denotes the dimensionless 21cm power spectrum, differentiated with respect to the parameters $p_i$, and $C$ is the data covariance matrix at redshift $z$ and scale $k$ for our error modelling given by Eq.~\eqref{eq:sigma21} and described in section~\ref{sec:21cmcq}. For reliability only the linear scale to each redshift is adopted for our analysis.

\begin{figure*}
  \includegraphics[width=1.0\textwidth]{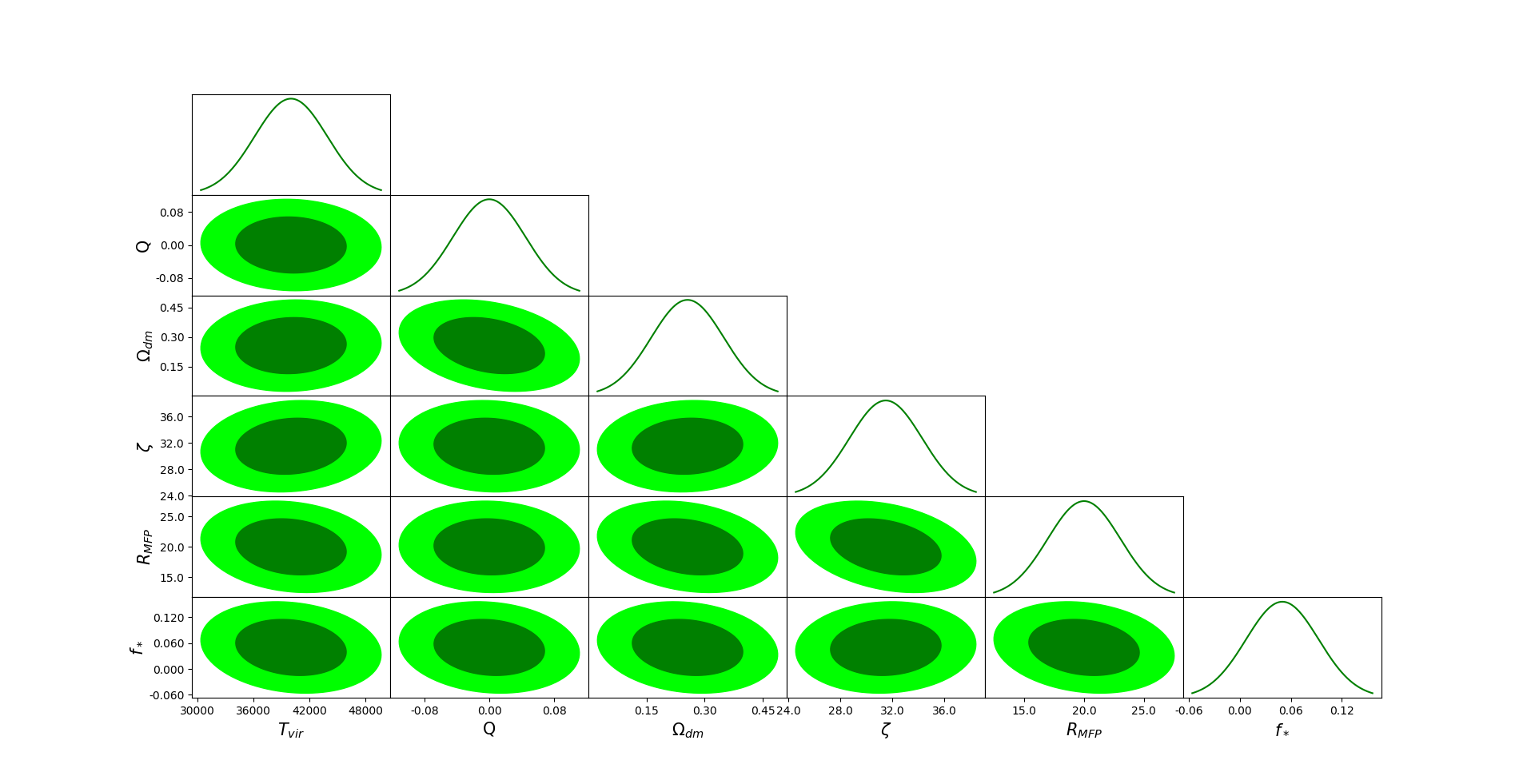}
  \caption{68.3\% and 95.4\% confidence ellipses derived by Fisher forecasts of 21cm power spectrum measurements, in the redshift range from $z=11$ to $z=6$. The fiducial value for parameters ($T_\mathrm{vir}, Q, \Omega_\mathrm{dm}, R_\mathrm{MFP}, \zeta, f_*$) are set as ($4\times 10^4, 0.0, 0.256, 20, 31.5, 0.05$) and $\lambda$ is chosen as 1.}\label{fig:2Dcounterpow}
\end{figure*}

Figure~\ref{fig:2Dcounterpow} shows the confidence contours derived for parameter set $\{ T_{\rm vir}, Q, \Omega_{\rm dm}, R_{\rm MFP}, \zeta, f_*\}$ for the combination of 21cm power spectra measured in six redshift bins from $z=6$ to $z=11$. We here now varied as astrophysical parameters the virial temperature $T_{\rm vir}$, the mean free path $R_{\rm MFP}$ and the ionization efficiency $\zeta$ that 21cm power spectra are usually sensitive to in the reionization model employed, as well as the fraction of baryons in stars $f_*$. The resulting marginalised $1\sigma$ confidence limits are shown in Table~\ref{tab:1sigmapmspow}. We note that the 21cm power spectrum measurements provide approximately the same constraints on the coupling~$Q$ as the global signal observation. Moreover, the power spectrum measurements are able to constrain the reionization parameters~$\zeta$ and $~R_\mathrm{MFP}$ within 10\% of the fiducial value, while simultaneously constraining cosmology. As compared to the global signal constraints, the resulting parameter constraints for the 21cm power spectra are almost independent of the fiducial value of the parameter $\lambda$ for the quintessence potential, as can be seen in Table~\ref{tab:1sigmapmspow}, since, as already noticed, the perturbation equation is independent of $\lambda$. The constraint on  $\Omega_\mathrm{dm}$ from 21cm global signal measurement is around $\mathcal{O}(0.005)$, which is one order of magnitude better than from the 21cm power spectrum and at the same level as provided by the Planck survey ~\citep{Ade:2015xua}.

    \begin{table}
    \centering
    \begin{tabular}{|c | c| c| c| c| c|c |}
     \hline
~ & $T_\mathrm{vir}$(K) & $Q$ & $\Omega_\mathrm{dm}$ & $\log(\zeta_\mathrm{X})$ & $f_{*}$  \\
 \hline
fiducial & $4\times 10^{4}$  & 0.0 & 0.256 & 129.638& 0.05 \\
 \hline
$\lambda$=1.0 & $222.71(0.56\%)$  & 0.022 & 0.0055(2.1\%) & 0.22(0.169\%) & 0.0034(6.8\%) \\
 \hline
$\lambda$=0.1 & $118.59(0.29\%)$  & 0.098 & 0.0043(1.7\%) & 0.097(0.075\%) & 0.0036(7.2\%) \\
 \hline
         \end{tabular}
\caption{
Marginalised 68.3\% confidence limits for parameter constraints derived from the global 21cm signal. The listed values are the   error and, in parentheses, the relative percentage error.}
    \label{tab:1sigmapmsglob}
    \end{table}

    \begin{table}
    \centering
    \begin{tabular}{| c| c| c| c| c| c|c |}
     \hline
 $\nu_\mathrm{res}$  &  $l_\mathrm{max}$ & $T_\mathrm{sys}$ & $t_\mathrm{int}$ & B (z=8) & $A_\mathrm{e}$ (z=8) & $n_{\perp}$  \\
    (kHz) & (cm) & (K) & (hrs) & (MHz) & (m$^2$) & \\
\hline
 3.9 &  $10^5$  & 400 & 1000 & 8  & 925 & 0.8 \\
 \hline
         \end{tabular}
\caption{Instrument specifications for the SKA stage 1 instrument as our fiducial 21cm experiment. Given are the spectral resolution $\nu_\mathrm{res}$, the maximum baseline $l_\mathrm{max}$, the instrument system temperature $T_\mathrm{sys}$, the total observing time time $t_\mathrm{int}$, the survey bandwidth $B$, the effective area $A_\mathrm{e}$ and the average number density of baselines $n\left( k_{\perp}\right)$ for mode $k_{\perp}$ perpendicular to the line-of-sight.}
    \label{tab:exp}
    \end{table}

    \begin{table}
    \centering
    \begin{tabular}{|c | c| c| c| c| c| c|c |}
     \hline
 ~ & $T_\mathrm{vir}$(K)  &  $Q$  &  $\Omega_\mathrm{dm}$  &  $\zeta$ &  $R_\mathrm{MFP}$(Mpc)  &  $f_{*}$  \\
 \hline
 fiducial & 4$\times 10^{4}$ &  $0.0$  & 0.256 & 20.0 & 31.5& 0.05 \\
 \hline
$\lambda=1.0$ & 3897.1(9.7\%) &  $0.044$  & 0.094(36.6\%) & 2.79(13.9\%) & 3.04(9.6\%) & 0.043(85.9\%) \\
 \hline
$\lambda=0.1$ & 2800.2(7.0\%) &  $0.036$  & 0.099(38.6\%) & 2.76(13.8\%) & 3.58(11.4\%)& 0.039(77.9\%) \\
 \hline
         \end{tabular}
\caption{Marginalised 68.3\% confidence limits for parameter constraints derived from 21cm power spectra. The listed values are the  error and, in parentheses, the relative percentage error. }
    \label{tab:1sigmapmspow}
    \end{table}

\section{Conclusion}\label{sec:conc}

The 21cm signal is a promising tool to probe the Universe from the dark ages to the formation of the first stars and to the end of reionization. In this paper we investigate the possibility to use the 21cm signal to constrain the coupling between dark matter and dark energy. This coupling scenario, which changes both the matter growth and cosmic background expansion, is taken into account in the calculation of 21cm emission by using a semi-numerical simulation based on the \texttt{21cmFAST} code. More explicitly, we derive the 21cm global signal between $z=15$ and $20$ and 21cm power spectra between $z=6$ and $11$ from our simulations, which covers the era when the first stars and galaxies emerge and when the Universe reionizes. Then we check the ability to detect a dark matter--dark energy coupling for both the 21cm global signal and fluctuation described by power spectra.

Using Fisher matrix forecasting, the result indicates that global signal measurements can improve the constraint on the dark matter density parameter $\Omega_\mathrm{dm}$ up to one order of magnitude in comparison with 21cm power spectra experiments. For the coupling constant Q  we can reach a precision up to the $\mathcal{O}(0.04)$ level both from 21cm global signal and power spectra, comparable with CMB results. On the astrophysical side, the two probes yield a similar level of constraint on $R_\mathrm{MFP}$ and $\zeta$ as well as a weaker one for $T_\mathrm{vir}$ compared to previous results \citep{Heneka:2018ins}. One interesting possibility for future studies would be to investigate constraints attainable on the  potential exponent  $\lambda$, studying as well its behaviour and degeneracies with respect to astrophysics. We opted in this first study to focus on the coupling, supported by our finding that our parameter forecast only mildly changes when varying $\lambda$ for a range of values suggested by previous constraints.

In summary, 21cm global signal detectors and  21cm power spectrum measurements are complementary and together reach a precision on some cosmological parameters comparable to  CMB experiments. The two probes bring forward plenty of information on both astrophysics and cosmology in a so far unexplored range of redshifts.

\section*{Acknowledgements}
Xue-Wen Liu would like to thank the China Scholarship Council, from whom he got the scholarship to stay one year in Heidelberg, Germany. He is also grateful to Dr. Wei-Ming Dai from University of KwaZulu-Natal in South Africa for the useful discussion.





\bibliographystyle{JHEP}
\bibliography{Coupledq}


\label{lastpage}
\end{document}